\newcommand{\std}[1]{
  \usepackage{amssymb}
  \usepackage{amsmath}
  \usepackage[final]{graphicx}
  \usepackage{a4,a4wide}
  \renewcommand{\baselinestretch}{#1}
  \parindent 0pt
  \parskip 2ex
  \pagestyle{headings}
  \begin{document}
}
\newcommand{\article}[1]{
  \documentclass[12pt,fleqn]{article}\std{#1}
}
\newcommand{\artikel}[1]{
  \documentclass[12pt,twoside,fleqn]{article}\usepackage{german}\std{#1}
}
\newcommand{\revtex}{
  \documentstyle[preprint,aps,eqsecnum,amssymb,fleqn]{revtex}
  \begin{document}
}
\newcommand{\book}[1]{
  \documentclass[12pt,twoside,fleqn]{book}
  \addtolength{\oddsidemargin}{2mm}
  \addtolength{\evensidemargin}{-4mm}
  \std{#1}
}
\newcommand{\buch}[1]{
  \documentclass[10pt,twoside,fleqn]{book}
  \usepackage{german}
  \usepackage{makeidx}
  \makeindex
  \std{#1}
}
\newcommand{\foils}[1]{
  \documentclass[fleqn]{article}
  \usepackages        
  \renewcommand{\baselinestretch}{#1}
  \setlength{\hoffset}{-4cm}
  \setlength{\voffset}{-4cm}
  \setlength{\textwidth}{20cm}
  \setlength{\textheight}{29cm}
  \parindent 0pt
  \parskip 4ex %8ex
  \pagestyle{empty}
  \begin{document}
  \Huge
}
\newcommand{\landfoils}[1]{
  \documentclass[fleqn]{article}
  \usepackage{amssymb}
  \usepackage{amsmath}
  \usepackage[final]{graphicx}
  \renewcommand{\baselinestretch}{#1}
  \setlength{\hoffset}{-5cm}%{-6cm}
  \setlength{\voffset}{-1.5cm}
  \setlength{\textwidth}{27cm}%{29cm}
  \setlength{\textheight}{19cm}
  \parindent 0ex
  \parskip 0ex %8ex
  \pagestyle{plain}
  \begin{document}
  \huge
}

\newcommand{\headline}[1]{\footnotetext{\sc Marc Toussaint, \today
    \hspace{\fill} file: #1}}

\newcommand{\sepline}{
\begin{center} \begin{picture}(200,0)
  \line(1,0){200}
\end{picture}\end{center}
}

\newcommand{\intro}[1]{\textbf{#1}\index{#1}}

\newtheorem{defi}{Definition}
\newenvironment{definition}{
\begin{quote}
\begin{defi}
}{
\end{defi}
\end{quote}
}

\newenvironment{block}{
\begin{quote} \begin{picture}(0,0)
        \put(-5,0){\line(1,0){20}}
        \put(-5,0){\line(0,-1){20}}
\end{picture}

}{

\begin{picture}(0,0)
        \put(-5,5){\line(1,0){20}}
        \put(-5,5){\line(0,1){20}}
\end{picture} \end{quote}
}

\newenvironment{summary}{
\begin{center}\begin{tabular}{|l|}
\hline
}{\\
\hline
\end{tabular}\end{center}
}

\newcommand{\inputReduce}[1]{
  
  {\sc\hspace{\fill} REDUCE file: #1}
%  \input{#1.tex}
%  \sepline
%  \input{../tex/tridefs}
%  \input{#1.out}
%  \redefinemath
}
\newcommand{\inputReduceInput}[1]{
  
  {\sc\hspace{\fill} REDUCE input - file: #1}
  \input{#1.tex}
}
\newcommand{\inputReduceOutput}[1]{
  
  {\sc\hspace{\fill} REDUCE output - file: #1}
%  \input{../tex/tridefs}
%  \input{#1.out}
%  \redefinemath
}

\newcommand{\macros}{
  \newcommand{\0}{{\hat 0}}
  \newcommand{\1}{{\hat 1}}
  \newcommand{\2}{{\hat 2}}
  \newcommand{\3}{{\hat 3}}
  \newcommand{\5}{{\hat 5}}
  \newcommand{\QQ}{{\cal Q}}

  \renewcommand{\a}{\alpha}
  \renewcommand{\b}{\beta}
  \renewcommand{\c}{\gamma}
  \renewcommand{\d}{\delta}
    \newcommand{\D}{\Delta}
    \newcommand{\e}{\epsilon}
    \newcommand{\g}{\gamma}
    \newcommand{\G}{\Gamma}
  \renewcommand{\l}{\lambda}
  \renewcommand{\L}{\Lambda}
    \newcommand{\m}{\mu}
    \newcommand{\n}{\nu}
    \newcommand{\N}{\nabla}
  \renewcommand{\k}{\kappa}
  \renewcommand{\o}{\omega}
  \renewcommand{\O}{\Omega}
    \newcommand{\p}{\varphi}
  \renewcommand{\r}{\varrho}
    \newcommand{\s}{\sigma}
    \newcommand{\Si}{\Sigma}
  \renewcommand{\t}{\theta}
  \renewcommand{\v}{\vartheta}
    \newcommand{\Y}{\Upsilon}

  \newcommand{\C}{{\bf C}}
  \newcommand{\R}{{\bf R}}
  \newcommand{\T}{{\bf T}}
  \newcommand{\Z}{{\bf Z}}

  \renewcommand{\AA}{{\cal A}}
  \newcommand{\GG}{{\cal G}}
  \newcommand{\TT}{{\cal T}}
  \newcommand{\EE}{{\cal E}}
  \newcommand{\HH}{{\cal H}}
  \newcommand{\II}{{\cal I}}
  \newcommand{\KK}{{\cal K}}
  \newcommand{\MM}{{\cal M}}
  \newcommand{\CC}{{\cal C}}
  \newcommand{\PP}{{\cal P}}
  \newcommand{\RR}{{\cal R}}
  \newcommand{\YY}{{\cal Y}}
  \newcommand{\SOSO}{{\cal SO}}
  \newcommand{\GLGL}{{\cal GL}}

  \newcommand{\ZZZ}{\mathbb{Z}}
  \newcommand{\RRR}{\mathrm{I\!R}}
  \newcommand{\CCC}{\mathbb{C}}

  \newcommand{\<}{\langle}
  \renewcommand{\>}{\rangle}
  \newcommand{\tr}{{\rm tr}}
  \newcommand{\lag}{\mathcal{L}}
  \newcommand{\inn}{\rfloor}
  \newcommand{\lie}{\pounds}
  \newcommand{\speer}{\parbox{0.4ex}{\raisebox{0.8ex}{$\nearrow$}}}
  \renewcommand{\dag}{ {}^\dagger }
  \newcommand{\h}{{}^\star}
  \newcommand{\w}{\wedge}
  \newcommand{\ow}{\stackrel{\circ}\wedge}
  \newcommand{\feed}{\nonumber \\}
  \newcommand{\comma}{\; , \quad}
  \newcommand{\period}{\; . \quad}
  \newcommand{\del}{\partial}
  \newcommand{\point}{$\bullet~~$}
  \newcommand{\doubletilde}{
  ~ \raisebox{0.3ex}{$\widetilde {}$} \raisebox{0.6ex}{$\widetilde {}$} \!\!
  }
  \newcommand{\topcirc}{\parbox{0ex}{~\raisebox{2.5ex}{${}^\circ$}}}
  \newcommand{\sym}{\topcirc}

  \newcommand{\half}{\frac{1}{2}}
  \newcommand{\third}{\frac{1}{3}}
  \newcommand{\fourth}{\frac{1}{4}}

}

\newcommand{\tmp}{\fbox{?}}
\newcommand{\Label}[1]{\label{#1}\fbox{\tiny #1}}

\macros
\newcommand{\path}{./}
\newcommand{\basepath}{./}
\newcommand{\setpath}[1]{
  \renewcommand{\path}{#1}
  \renewcommand{\basepath}{#1}}
\newcommand{\pathinput}[2]{
  \renewcommand{\path}{\basepath #1}
  \input{\path #2} \renewcommand{\path}{\basepath}}
\article{1}
\renewcommand{\Label}[1]{\label{#1}}

\title{A numeric solution for metric-affine gravity and Einstein's gravitational theory with Proca matter}
\author{
Marc Toussaint\\
\small Institute for Theoretical Physics, University of Cologne\\
\small 50923 K\"oln, Germany\\
\small {\tt www.thp.uni-koeln.de/\~{}mt/}}

\maketitle

\begin{abstract}
  A special case of metric-affine gauge theory of gravity (MAG) is
  equivalent to general relativity with Proca matter as source. We
  study in detail a corresponding numeric solution of the
  Reissner-Nordstr\"om type. It is static, spherically symmetric, and
  of electric type. In particular, this solution has no horizon, so it
  has a naked singularity as its origin.
\end{abstract}

\newpage \parskip 0ex
\tableofcontents
\parskip 2ex

\section{Introduction}

We will present a numeric solution of the Einstein-Proca theory with
motivation that this theory is equivalent to a special case of the
metric-affine gauge theory of gravity (MAG) \cite{hehlMMN}. This
special case is the so-called triplet ansatz \cite{obuVEH} in which,
roughly, the lagrangian includes square terms of nonmetricity but also
square terms of the derivative of nonmetricity (stemming from
curvature squares). It is a special feature of this ansatz that the
nonmetricity (as well as torsion) may be expressed by a 1-form, which,
because of this type of lagrangian, represents a massive, propagating
1-form field, i.e.\ perfectly equivalent to a Proca field. Obukhov et
al.\ \cite{obuVEH} gave an exact proof of this equivalence.

The outline of this paper is rather straightforward: First, we briefly
present the triplet ansatz and Obukhov's theorem. Then we review the
Einstein-Proca theory, set up the lagrangian, derive the field
equations, and simplify them for a spherically symmetric ansatz. After
addressing the problem of integration constants and dimensions we can
perform the numerical integration. We also present a power series
expansion of the solution around the origin. The most important
features of our solution are summarized at the end.

\subsubsection*{MAG and the triplet ansatz}

Given the curvature $R_\a{}^\b$ and its 6+5 irreducible pieces
${}^{(I)}\hat R_\a{}^\b$ and ${}^{(I)}\sym R_\a{}^\b$, the torsion
$T_\a$ and its 3 irreducible pieces ${}^{(I)}T_\a$, and the
nonmetricity $Q_{\a\b}$ and its 4 irreducible pieces
${}^{(I)}Q_{\a\b}$ (see \cite{hehlMMN} appendix B), we may write the
general MAG lagrangian as
\begin{align}
\lag_{\rm MAG} =~ \frac{1}{2\kappa}\Big[
& -~ a_0 R^{\a\b} \w \eta_{\a\b} - 2 \l\eta
                                          &&\mbox{\`a la Hilbert-Einstein}\feed
& +~ T^\a \w\h (a_{I=1..3} {}^{(I)}T{}_\a)
                                          &&\mbox{quadratic torsion}\feed
& +~ Q_{\a\b} \w\h (b_{I=1..4} {}^{(I)}Q^{\a\b})
                                          &&\mbox{quadratic nonmetricity} \feed
& +~ b_5\, ({}^{(3)}Q_{\a\g} \w \v^\a) \w\h ({}^{(4)}Q^{\b\g} \w \v_\b)
                                          &&\mbox{quad.\ nonm.\ mixed with $\v^\a$}\feed
& +~ 2 (c_{I=2..4} ~^{(I)}Q_{\a\b}) \w \v^\a \w \h T^\b \Big]
                                          &&\mbox{cross terms nonm./torsion}\feed
& \hspace{-1.2cm} -~ \frac{1}{2\r} R^{\a\b} \w\h \Big[ 
     (w_{I=1..6} {}^{(I)}\hat R_{\a\b}) + (z_{I=1..5} {}^{(I)}\sym\! R_{\a\b})
                                          &&\mbox{quadratic curvature}\feed
&    + w_7\,        \v_\a \w (e_\g \inn   {}^{(5)}\hat R^\g{}_\b)
     + z_6\,        \v_\g \w (e_\a \inn   {}^{(2)}\sym\! R^\g{}_\b) &&\feed
&    + z_{I=7..9}\, \v_\a \w (e_\g \inn {}^{(I-4)}\sym\! R^\g{}_\b) \Big] \period
                                          &&\raisebox{1ex}{$\Big\}$mixed with $\v^\a$}
\Label{generalMAG}
\end{align}
Here, we sum over the repeated index $I$. This lagrangian and the
presently known solutions have been reviewed in \cite{hehlM}. We have
the \emph{weak} and \emph{strong} coupling constants $1/\k$ and
$1/\r$, the cosmological constant $\l$, and the 28 parameters
$a_{I=0..3}$, $b_{I=1..5}$, $c_{I=2..4}$, $w_{I=1..7}$, and
$z_{I=1..9}$. Note that the weak coupling constant has length
dimension $[1/\k]=1/\ell^2$ because it multiplies to a torsion square,
torsion being the field strength of translation generators with
dimension $1/\ell$. The strong coupling constant, though, has no
length dimension. The work presented here is only concerned with the
so-called triplet ansatz, i.e.\ the special case
\begin{align}
\r=1 \comma w_{I=1..7}=0 \comma z_{I=1..3}=z_{I=5..9}=0 \comma z_4 \not =0
\period
\Label{lagconstraint}
\end{align}
This means that we consider a general \emph{weak lagrangian} (with
weak coupling constant) but only a very restricted \emph{strong
  lagrangian} (of curvature squares) allowing only for a square of the
dilation curvature ${}^{(4)}\sym R_{\a\b}:=\frac{1}{n}\, g_{\a\b}\,
R_\g{}^\g$. Qualitatively, the lagrangian with these constraints may
be displayed as
\begin{align}
\lag ~~\sim~~ \l ~+~ R ~+~ \big(T+Q\big)^2 ~+~ \big({}^{(4)}\sym R\big)^2 \;.
\Label{lag2}
\end{align}
Here, $R$, $T$, and $Q$ denote just some terms linear in the
curvature, torsion, and nonmetricity, respectively. On this
qualitative level, the result of Obukhov et al.\ \cite{obuVEH} is the
following: Effectively, the curvature $R$ may be considered
Riemannian, $T$ and $Q$ may be replaced by a 1-form $\phi$, i.e.\ 
$T + Q \sim \phi$, and ${}^{(4)}\sym\! R$ is similar to $d\phi$. Hence,
(\ref{lag2}) reads generically
\begin{align}
\lag ~~\sim~~ \l ~+~ R_{\rm riem} ~+~ \phi^2 ~+~ (d\phi)^2 \;,
\Label{quallag}
\end{align}
which describes an Einstein spacetime with a massive 1-form field $\phi$,
i.e.\ a Proca field.

We will review the results of \cite{obuVEH} in detail. First, one
considers \emph{a special case} of the MAG lagrangian
(\ref{generalMAG}) with constraint (\ref{lagconstraint}) by specifying
the remaining parameters $\l$, $a_{I=0..3}$, $b_{I=1..5}$,
$c_{I=2..4}$, and $z_4$.  This choice is done in \cite{obuVEH} eq
(4.1) and turns out to effectively produce a purely Riemannian
Hilbert-Einstein lagrangian, cf.\ \cite{obuVEH} (4.6). This allows to
introduce a new variable, the effective Riemannian curvature. Then,
having investigated this special lagrangian, they generalize it again
by adding a general lagrangian (restricted by (\ref{lagconstraint}))
to it, \cite{obuVEH} (5.1-5.5).  With the aid of the effective
Riemannian curvature, the field equation FIRST (the variation with
respect to the anholonomic coframe, see \cite{hehlMMN}) reads like an
Einstein equation with an energy-momentum source that depends on
torsion and nonmetricity, \cite{obuVEH} (5.11). The field equation
SECOND (the variation with respect to the linear connection) becomes a
system of differential equations for torsion and nonmetricity alone.
In the vacuum case, where the energy-momentum $\Si_\a$ and the
hypermomentum $\D_\a{}^\b$ of matter vanish, these differential
equations (i.e.\ SECOND) reduce to
\begin{align}
& ^{(1)}T_\a = 0 \comma
  ^{(2)}T_\a = \frac{k_2}{3}\, \v_\a \w \phi \comma
  ^{(3)}T_\a = 0 \comma
&&\mbox{cf.\ \cite{obuVEH} (6.2,6.6+2.3,5.20)}\feed
& ^{(1)}Q_{\a\b} = 0 \comma
  ^{(2)}Q_{\a\b} = 0 \comma
&&\mbox{cf.\ \cite{obuVEH} (5.27,6.3)}\feed
& ^{(3)}Q_{\a\b} = \frac{4}{9} k_1\, \Big(\v_{(\a} e_{\b)}\inn \phi - \frac{1}{4} g_{\a\b}\, \phi\Big) \comma
  ^{(4)}Q_{\a\b} = k_0\, g_{\a\b}\, \phi \comma
&&\mbox{cf.\ \cite{obuVEH} (6.5+2.7,2.8)}\feed
& d \h d \phi + m^2 \phi = 0 \period
&&\mbox{cf.\ \cite{obuVEH} (6.7)}
\Label{triplet}
\end{align}
Here, the new 1-form $\phi$ determines the torsion and nonmetricity
completely and needs to solve the Proca equation. The four constants
$m$, $k_0$, $k_1$, and $k_2$ uniquely depend on the parameters of the
MAG lagrangian via \cite{obuVEH} (6.8, 6.4, 5.3-5.5).  We summarize
\begin{align}
k_0 &= 4 (a_2 - 2a_0)(b_3 + a_0/8) - 3 (c_3 + a_0)^2 \comma\feed
k_1 &= 9/2 (a_2 - 2a_0)(b_5 - a_0) - 9 (c_3 + a_0)(c_4 + a_0) \comma\feed
k_2 &= 12 (b_3 + a_0/8)(c_4 + a_0) - 9/2 (b_5 - a_0)(c_3 + a_0) \comma\feed
m^2 &= \frac{1}{z_4 \k}\left(-4 b_4 + \frac{3}{2}\, a_0 
     + \frac{k_1}{2k_0}\, (b_5 - a_0) + \frac{k_2}{k_0}\, (c_4 + a_0)\right)
\period
\Label{constants}
\end{align}
Obviously, these parameters depend only on $a_0$, $a_2$, $b_3$, $b_4$,
$b_5$, $c_3$, $c_4$, and $z_4$. With (\ref{triplet}) and
(\ref{constants}) we can express the energy-momentum source of torsion
and nonmetricity in the effective Einstein equation in terms of $\phi$
\cite{obuVEH} (7.3, 7.5). This energy-momentum is exactly the
energy-momentum of the Proca 1-form $\phi$. Hence, we finally found
that the MAG lagrangian (\ref{generalMAG}), restricted by
(\ref{lagconstraint}), together with its field equations, is
effectively equivalent to an Einstein-Proca lagrangian as suggested in
(\ref{quallag}). The parameter $m$ given in (\ref{constants}) has the
meaning of the mass parameter of the Proca 1-form $\phi$. If $m$
vanishes, the initial MAG theory is equivalent to the Einstein-Maxwell
theory. In general, $m=0$ is equivalent to
\begin{align}
0&=
 32\,{b_{4}}\,{a_{2}}\,{b_{3}} - 12\,{a_{0}}\,{a_{2}}\,{b_{3}}
- 64\,{b_{4}}\,{a_{0}}\,{b_{3}} - 24\,{b_{3}}\,{c_{4}}^{2}
- 48\,{b_{3}}\,{c_{4}}\,{a_{0}}
- 32\,{b_{4}}\,{a_{0}}^{2} - 24\,{b_{4}}\,{c_{3}}^{2}\feed
&
+ 9\,{a_{2}}\,{b_{5}}\,{a_{0}} - 6\,{a_{2}}\,{a_{0}}^{2}
+ 9\,{a_{0}}\,{c_{3}}^{2} - 48\,{b_{4}}\,{c_{3}}\,{a_{0}}
+ 4\,{b_{4}}\,{a_{2}}\,{a_{0}} - 24\,{a_{0}}^{2}\,{c_{4}}
+ 9\,{a_{0}}\,{b_{5}}^{2} \feed
&
- \frac {9}{2} \,{a_{2}}\,{b_{5}}^{2} - 3\,{a_{0}}\,{c_{4}}^{2}
+ 18\,{c_{3}}\,{c_{4}}\,{b_{5}} - 18\,{c_{3}}\,{c_{4}}\,{a_{0}}
+ 18\,{c_{3}}\,{a_{0}}\,{b_{5}} + 18\,{a_{0}}\,{c_{4}}\,{b_{5}}
\period
\Label{zeromass}
\end{align}
This equation generalizes \cite{obuVETH} (4.2) for $b_5 \not =0$.
Thus, the exact solution found in \cite{obuVETH} with $m=0$
corresponds to an exact solution of an Einstein-Maxwell system. Here,
we want to present a solution for $m \not = 0$.

\section{The Einstein-Proca theory}

Motivated by the previous section we now concentrate on the Proca
lagrangian $\lag_P$ of a massive 1-form $\phi$:
\begin{align}
\lag_P&=
- \frac{1}{2}\, d\phi \w\h d\phi + \frac{1}{2}\, m^2 \phi \w\h \phi \period
\Label{procalag}
\end{align}
First, we shortly discuss the dimension of $m$. We know that the
Hodge-dual of a $p$-form is an $(n\!-\!p)$-form. Hence, whenever the
Hodge-dual applies on a $p$-form, it has the dimension
$[\,\h\,]=\ell^{n-2p}$. It follows that $[d\phi \w\h d\phi]=[\phi]^2$,
and $\phi \w\h \phi=\ell^2 [\phi]^2$. To be able to consistently add
these terms in the lagrangian the dimension of the mass parameter
needs to be $[m]=1/\ell$.

It is straightforward to calculate the Proca field equation and the
canonical energy-momentum of this lagrangian with the Noether-Lagrange
machinery presented in \cite{hehlMMN}. The variations yield:
\begin{align}
\frac{\d \lag_P}{\d \phi}
&= d \frac{\del \lag_P}{\del (d\phi)} + \frac{\del \lag_P}{\del \phi}
= - d \h d \phi + m^2\, \h \phi\comma \\
\frac{\delta \lag_P}{\delta \v^\a}
&=: \Si_\a
= e_\a \inn \lag_P
- (e_\a \inn d\phi) \frac{\del \lag_P}{\del d\phi}
- (e_\a \inn \phi) \frac{\del \lag_P}{\del \phi} \feed
&=\half\left[-(e_\a \inn d\phi)\w\h d\phi - d\phi \w (e_\a \inn \h d\phi)
+ m^2 (e_\a \inn \phi) \w\h\phi - m^2 \phi \w (e_\a \inn \h \phi)\right]\feed
&\quad + (e_\a \inn d\phi)\w\h d\phi - m^2(e_\a \inn \phi) \w \h \phi\feed
&=\half\left[(e_\a \inn d \phi) \w \h d \phi - d\phi \w  (e_\a \inn \h d \phi)
- m^2 \left[ (e_\a \inn \phi) \w \h \phi + \phi \w (e_\a \inn \h \phi) \right]
\right]
\period
\Label{procaem}
\end{align}
Coupled with a Riemannian background, i.e.\ considering a lagrangian\\
$\lag=\lag_P + \lag_{\rm Hilbert-Einstein}$, we end up with the field
equations
\begin{align}
&0 = - d\h d \phi + m^2\, \h \phi \comma && \mbox{Proca equation} \comma \Label{procaeq}\\
&0 = G_\a - \k \Si_\a =: X_\a \comma &&\mbox{Einstein equation}  \period \Label{einsteineq}
\end{align}
For completeness we also display the contracted Bianchi identities
\begin{align}
&0 = d \Si^\a + \G_\b{}^\a \w \Si^\b \;.\Label{bianchieq}
\end{align}
Also Obukhov and Vlachynsky \cite{obuV} considered this system and
indeed found the same solution we will find.  Unfortunately, they did
not publish their results until recently such that the author did not
know about their efforts for a long time. However, the integration and
the results are presented here in much more detail. Before we
concentrate on a solution of this system, we consider two modified
versions of the problem.

\subsection{A Proca field on flat spacetime background}\Label{Cflat}

If we consider the lagrangian (\ref{procalag}) on \emph{flat}, i.e.\ 
Minkowskian background we find two simple solutions. The first we find
by an ansatz in analogy to the static \emph{electric} monopole field
$A_{\rm el}=q/r\, dt$: We suppose $\phi$ to be a static, spherically
symmetric, and \emph{time-like} 1-form
\begin{align}
\phi=\frac{u_{\rm el}(r)}{r}\, dt \;,
\end{align}
where $t$ denotes the time coordinate and $r$ the radius in spherical
coordinates. With this ansatz, the Proca equation (\ref{procaeq})
becomes
\begin{align}
0 = \frac{1}{r}\, (-u_{\rm el}'' + m^2\, u_{\rm el})
\Label{procaflateq}
\end{align}
and can be solved by the well known Yukawa potential
\begin{align}
\frac{u_{\rm el}(r)}{r}=\frac{q}{r}\,\exp(-mr) \;.
\end{align}
The parameter $q$ is called Proca charge. A second solution we find
by considering an ansatz in analogy to the static \emph{magnetic}
monopole field $A_{\rm mag}=p\,(1-\cos\t)\, d\p$: We set
\begin{align}
\phi=u_{\rm mag}(r)\,(1-\cos\t)\, d\p \;.
\end{align}
The field equations (\ref{procaeq}) becomes
\begin{align}
0 = \frac{1}{r\,\sin\t}\, (1-\cos\t)(-u_{\rm mag}'' + m^2\,u_{\rm mag})
\end{align}
which is, of course, also solved by $u_{\rm mag}=p\,\exp(-mr)$, where
$p$ might be called \emph{magnetic} Proca charge.

\subsection{A Proca vector field}

Here we want to clarify that there \emph{is} a difference between an
ansatz of $\phi$ as a 1-form and as a vector. Consider the analogy of
the lagrangian (\ref{procalag}) for a vector-valued 0-form $\phi_\a$:
\begin{align}
\lag_P
&=- D\phi^\a \w \h D\phi_\a  \; + \;  m^2 \phi^\a \w \h \phi_\a \comma\\
\frac{\d \lag_P}{\d \phi^\a}
&= - D \frac{\del \lag_P}{\del (D\phi^\a)} + \frac{\del \lag_P}{\del \phi^\a}
= + 2\, D \h D \phi_\a + 2\, m^2\, \h \phi_\a \comma\Label{procaveceq}\\
\frac{\delta \lag_P}{\delta \v^\a}
&=: \Si_\a
= e_\a \inn \lag_P
- (e_\a \inn D\phi^\b) \frac{\del \lag_P}{\del D\phi^\b}
- (e_\a \inn \phi^\b) \frac{\del \lag_P}{\del \phi^\b} \feed
&=-(e_\a \inn D\phi^\b)\w\h D\phi_\b + D\phi^\b \w (e_\a \inn \h D\phi_\b)
+ m^2 (e_\a \inn \phi^\b) \w\h\phi_\b + m^2 \phi^\b \w (e_\a \inn \h \phi_\b)\feed
&+ 2(e_\a \inn D\phi^\b)\w\h D\phi_\b - 2m^2(e_\a \inn \phi^\b) \w \h \phi_\b\feed
&=(e_\a \inn D\phi^\b) \w \h D\phi_\b + D\phi^\b \w  (e_\a \inn \h D\phi_\b)
- m^2 \left[ (e_\a \inn \phi^\b) \w \h \phi_\b - \phi^\b \w (e_\a \inn \h \phi_\b) \right]\feed
&=(e_\a \inn D\phi^\b) \w \h D\phi_\b + D\phi^\b \w  (e_\a \inn \h D\phi_\b)
+ m^2 \phi^\b \w (e_\a \inn \h \phi_\b)
\period\Label{procavecem}
\end{align}
Note that we introduced the covariant derivative $D=d+\G^{\{\}}$ with
the Levi-Civita connection $\G^{\{\}}$. In flat space, say, we find
that the signs in (\ref{procaveceq}) and (\ref{procaeq}) do not agree.
Also, the energy-momentum (\ref{procavecem}) is very different from
(\ref{procaem}). From the introduction it is clear that we are
interested in a Proca 1-form and not in a Proca vector field.

\section{An solution with electric type Proca field}

Motivated by the Yukawa solution $\phi=\frac{q\,\exp(-mr)}{r}\, dt$ of
the flat Proca equation and the analogy to the electric monopole, we
make the ansatz that $\phi$ is a spherically symmetric, static, and
time-like 1-form:
\begin{align}
\phi=\frac{u(r)}{r} dt \;.
\end{align}
The general spherically symmetric ansatz for the coframe (i.e.\ 
implicitly for the metric) reads
\begin{align}
\v^{\hat 0} = f\, dt \comma \v^{\hat 1} = \frac{g}{f}\, dr \comma
\v^{\hat 2} = r\, d\t   \comma \v^{\hat 3} = r \sin\theta\, d\p  \comma
f=f(r) \comma g=g(r) \period
\Label{coframe}
\end{align}
The meaning of the function $g$ is illustrated by the relation
$\eta=g\, \eta_{\rm flat}$ for the volume element $\eta = \h 1$.
Writing the Proca field as $\phi=\Phi(r)\, dt$, with $\Phi(r)
:=u(r)/r$, the energy-momentum (\ref{procaem}) of this ansatz turns
out to be
\begin{align}
  \v^{\hat 0} \w \Si^{\hat 0}&= \frac{\eta}{2} \left[ -(\Phi'/g)^2 -
    m^2 \Phi^2/f^2 \right] \comma\feed \v^{\hat 1} \w \Si^{\hat 1}&=
  \frac{\eta}{2} \left[ ~ ~ \,(\Phi'/g)^2 - m^2 \Phi^2/f^2 \right]
  \comma\feed \v^{\hat 2} \w \Si^{\hat 2}&= \frac{\eta}{2} \left[
    -(\Phi'/g)^2 - m^2 \Phi^2/f^2 \right] \comma\feed \v^{\hat 3} \w
  \Si^{\hat 3}&= \frac{\eta}{2} \left[ -(\Phi'/g)^2 - m^2 \Phi^2/f^2
  \right] \comma\feed \v^\m \w \Si^\n&=0 \quad \mbox{for $\m \not =
    \n$} \comma \feed \h ( \v^\a \w \Si_\a ) &= -\frac{m^2 \Phi^2}{f^2}
  \comma\feed \h ( \Si_\a \w\h \Si^\a ) &= \Big(\frac{\Phi'}{g}\Big)^4 +
  \Big(\frac{\Phi' m \Phi}{fg}\Big)^2 + \Big(\frac{m \Phi}{f}\Big)^4
  \period
\Label{energymomentum}
\end{align}
Recall that $\h(\v^\m \w \Si^\n) =: T^{\m\n}$ represent the components
of the ordinary 2nd rank energy-momentum tensor. The non-vanishing
components of equations (\ref{procaeq}), (\ref{einsteineq}), and
(\ref{bianchieq}) read in simplified form
\begin{align}
\text{P}_0:&~~ 0= f^2g'(u-ru') + gr(f^2u''-g^2m^2u) \feed
\text{P}_1:&~~ 0= g\dot u + \dot g(ru'-u) \feed
\text{E}_{00}:&~~ 0= -\k\big[(u-ru')^2 +m^2r^2u^2g^2/f^2\big]+2r^2(f^2-g^2) +4ff'r^3  -4f^2r^3g'/g \feed
\text{E}_{11}:&~~ 0= -\k\big[(u-ru')^2 -m^2r^2u^2g^2/f^2\big]+2r^2(f^2-g^2) +4ff'r^3  \feed
\text{E}_{01}:&~~ 0= f \dot g - g\dot f\feed
\text{E}_{22}:&~~ 0= -\k g f^4\big[(u-ru')^2 +m^2r^2u^2g^2/f^2\big] +2g^2r^4f(f \ddot g -g\ddot f) - 6\dot f (f\dot g-g\dot f)\feed &~~~~~ +2f^4r^4f'(fg'-gf') -2f^5f''gr^4 +2f^5r^3(fg'-2gf')\feed 
\text{B}_0:&~~ 0= f^3(ru'-u)\big[g\dot u +\dot g(ru'-u)\big] -g^2m^2r^2u \big[g(f\dot u -u\dot f) +u(f\dot g- g\dot f)\big]\feed
\text{B}_1:&~~ 0= f^2g'(u-ru') +gr(f^2u'' -g^2m^2u)
\Label{einsteinproca}
\end{align}
Here, P, E, and B refer to the Proca, Einstein, and contracted Bianchi
equations and indices denote the respective components. These
equations also include, for later purposes, the case of time dependent
functions $u$, $f$, and $g$. Here, we neglect this time dependence and
thus discard all time derivatives. We see that $\text{B}_1$ is
equivalent to $\text{P}_0$ and that $\text{E}_{11}-\text{E}_{00}$
simplifies the Einstein equations a lot. We end up with the following
ordinary differential equations system of second order in $u$ and
first order in $f$ and $g$:
\begin{align}
\text{P}_0:&~~ 0= f^2g'(u-ru') + gr(f^2u''-g^2m^2u)\;,\Label{ee1}\\
\text{E}_{11}:&~~ 0= -\k\big[(u-ru')^2 -m^2r^2u^2g^2/f^2\big]+2r^2(f^2-g^2) +4ff'r^3\;,\Label{ee2}\\
\text{E}_{11}-\text{E}_{00}:&~~ 0= \k m^2u^2g^3 + 4f^4rg'\;.\Label{ee3}
\end{align}
These equations are equivalent to \cite{obuV} eqs (3.5, 3.9, 3.10)
found by Obukhov and Vlachynsky. We did most of the calculations with
the aid of the computer algebra systems Reduce and Maple. The
respective files can be found at
\verb+www.thp.uni-koeln/~mt/work/1999diplom/+.

\subsection{Preparing the numerical integration}

First we briefly discuss the dimensions of the system. We know the
dimensions of the radius, the gravitational coupling constant, and the
Proca mass, $[r]=\ell$, $[\k]=\ell^2$, $[m]=1/\ell$, respectively.
Hence, we get rid of all dimensions by rescaling the radius variable
$r \to r/\sqrt{\k}$ and the mass parameter $m \to \sqrt{\k}\, m$. In
practice, i.e.\ when investigating the equation system with computer
algebra, we simply put $\k \equiv 1$, which is equivalent to the
rescaling but saves us from introducing new variables. Also, we can
eliminate the mass parameter $m$ from (\ref{ee1}-\ref{ee3}) by the
substitution $r \to m\,r$, $f\to f/m$, and $g\to g/m$. Instead, again,
we equivalently fix $m=1$ in the following without loosing generality.

We now concentrate on the dimensionless ordinary differential equation
system (\ref{ee1}-\ref{ee3}). Parameter $m$ is fixed and obviously
\emph{not} an integration constant. The equation system is of first
order in $f$ and $g$, and of second order in $u$. It is easy to reduce
it to an ordinary first order differential equation system by
substituting $u' \to v$, $u'' \to v'$ and by adding a fourth equation
$v=u'$ to the system. Hence, a general integration of this system
leads to four integration constants. In the case of a numerical
approach, these constant are the initial values for $f$, $g$, $u$, and
$u'$ at the starting point of integration.  Here, we consider only two
starting points $r_0 \ll 1$ and $r_\infty \gg 1$ for integrations from
zero and infinity. We denote the set of integration constants by
$(f_0, g_0, u_0, u'_0)$ in one case and by $(f_\infty, g_\infty,
u_\infty, u'_\infty)$ in the other. The discussion above suggests that
we will find a 4-parameter set of solutions. But this is misleading.
In principle it is possible to start integration with four arbitrary
parameters at some point $\tilde r \in \RRR$. But then, in general,
one will find a solution of the equations (\ref{ee1}-\ref{ee3}) only
in a neighborhood of $\tilde r$. More precisely, the convergence for $r \to 0$ and $r \to \infty$ are 2 constraints and thus we expect only a 2-parameter set of suitable integration constants.

First we discuss the limit $r \to 0$. We presume that $g$, $u$, and
$u'$ are finite at zero and that $f$ diverges as $1/r$. This can be
motivated by the fact that such an ansatz solves the system of
equations (\ref{ee1}-\ref{ee3}) as we will show shortly, or by the
arguments that $g$ needs to be finite because of the volume element,
$u$ and $u'$ should be finite because of the energy, and $f$ should
behave as $1/r$ because of the analogy to the Reissner-Nordstr\"om
solution.  Later, also our numerical integrations confirm this
assumption. To investigate the limit $r\to 0$ we therefore insert the
expansion
\begin{align}
u&=u_1 + u_2\,r + u_3\,r^2 +.. \comma\feed
g&=g_1 + g_2\,r + g_3\,r^2 +.. \comma\feed
f&=f_1/r + f_2 + f_3\,r +.. \Label{expansion}
\end{align}
into the system (\ref{ee1}-\ref{ee3}). Collecting the coefficients of
0-th order in $r$ in each of the three equations we find respectively
\begin{align}
0&=f_1\, g_2\, u_1 \;,\feed
0&=f_1^4\,g_1 - \half h_1^2\,g_1\,u_1^2 \;,\feed
0&=h_1^4\,g_2 \;.
\end{align}
These equations are solved by $g_2=0$ and $f_1=u_1/\sqrt{2}$. This
means that, for $r \to 0$, we find the following approximate solution
of the equation system (\ref{ee1}-\ref{ee3}):
\begin{align}
f_0 &= \frac{q}{\sqrt{2} r_0} \comma\feed
g_0 &= c \comma\feed
u_0 &= q \comma\feed
u'_0 &= b \period
\Label{zeroconst}
\end{align}
Hence, for an integration from zero we are left \emph{three}
parameters $(q,b,c)$. Our results in section \ref{Cfixedq} will show
that such solutions with arbitrary $(q,b,c)$ diverge at some finite
radius. Only a fine tuning of one of the parameters (we tuned $c$)
makes the solution converging at infinity.  Hence, regarding only
converging, global solutions, we are left with a 2-parameter set of
solutions parameterized by $(q,b)$.

For the limit $r_\infty$ it is natural to require the metric function
to tend to the Schwarzschild solution with the new mass parameter $M$
that represents the total gravitating mass. Also, the Proca field
should tend to the vacuum solution of the Proca equation, i.e.\ the
Yukawa potential specified by the new Proca charge parameter $Q$. In
detail we would expect
\begin{align}
f_\infty &= \sqrt{1-\frac{2M}{r_\infty}} \comma\feed
g_\infty &= 1 \comma\feed
u_\infty &= Q\, \exp(-m r_\infty) \comma\feed
u'_\infty &= -m Q\,  \exp(-m r_\infty) \period
\Label{inftyconst}
\end{align}
This gives a 2-parameter set of solutions parameterized by $(M,Q)$.
Since we sill find (numerically) that all of these solutions converge
for $r \to 0$ we conclude that this set of solutions contains all
global solutions.

\subsubsection*{A power series expansion}
There exists a simple scheme to determine all coefficients in the
expansion (\ref{expansion}): Inserting the expansion and considering
the coefficients of $i$-th order ($i\ge 2$) in $r$ in each of the
three equations (\ref{ee1}-\ref{ee3}), one can solve for $u_{i+2}$,
$g_{i+2}$, and $f_{i+1}$ in terms of $u_{j+2}$, $g_{j+2}$, and
$f_{j+1}$ with $j < i$. This iteration is very easily implemented in
Maple and we display here the result after considering the equations
up to 4-th order:
\begin{align}
u(r)=
&q + b\,r
- \frac {1}{10}\,\frac {c^2\,b}{q^2}\,r^5 \;,\\
g(r)=
&c + { \frac {1}{2}} \,\frac {c^{3}}{q^{2}}\,r^{4}
+ \frac {4}{5}\,\frac {c^3\,b}{q^3}\,r^5 \;,\\
f(r)=
&\frac {1}{\sqrt 2}\,\frac{q}{r} 
+ \frac {1}{\sqrt 2}\,\frac {c^2}{q}\,r 
+ \frac {1}{6 \sqrt 2}\,\frac {2\,q^2\,c^2 - 3\,c^4}{q^3}\,r^3 \;.
\Label{power}
\end{align}
The problems is that an insertion of the power expansion
(\ref{expansion}) into the equation system exceeds the computer's
memory resources very fast. Being limited in this way, we could not
observe an appropriate convergence behavior for large $r$.

\newcommand{\plot}[2]{
\includegraphics[scale=#2,angle=-90,bb=100 100 530 650]{proca_#1.eps}}

\subsection{Integration from infinity for various $M$ and $Q$}

We perform the numerical integration with the standard Runge-Kutta
method provided by the computer algebra system Maple. (In detail: We
used the {\tt rkf45} method with 15 digits, absolute ({\tt abserr})
and relative ({\tt relerr}) errors $10^{-13}$, and unlimited number of
function evaluations ({\tt maxfun}).) Such an integration takes only
about 10 seconds. You can find all calculations in the Maple-file
given.

\begin{figure}[t]
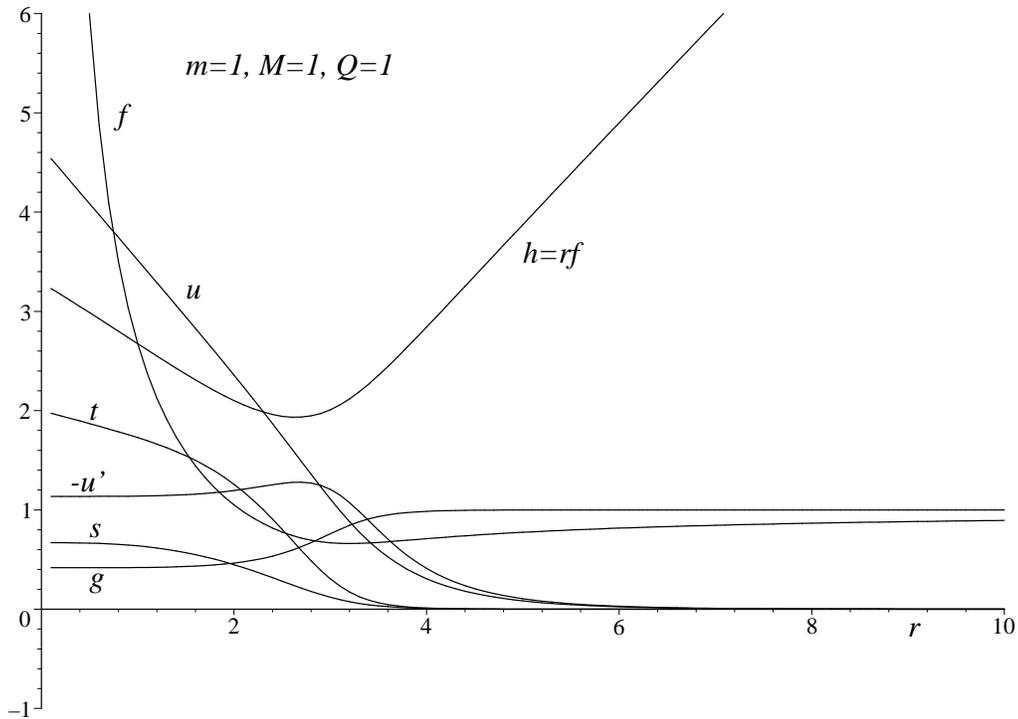
\center
\plot{all}{0.7}
\caption{
  Typical solution: Integration of (\ref{ee1}-\ref{ee3}) for $m=M=Q=1$
  from $r_\infty=60$, with integration constants (\ref{inftyconst})
  performed by Maple. We see the metric functions $f$, $g$, and
  $h:=r\,f$, the Proca function $u$ and $u'$, and the energy-momentum
  trace $t$ and its integral $s$ (divided by 100).}
\Label{Tall}
\end{figure}
Figure \ref{Tall} represents a typical solution for an integration from
infinity ($r_\infty=60$, which is far enough from the Schwarzschild
radius $r_{\rm S}(M)$). The gravitating mass $M$ and the Proca charge
$Q$ are of the same order as $m = 1$. We see the metric functions $f$
and $g$. To get a better impression of the behavior of $f$ for $r \to
0$ we add a plot of $h:=r f$. The asymptote of $h$ for $r \to \infty$
($h\sim r\sqrt{1-2M/r} \sim r-M$) crosses the r-axis at $r=M$. The
Proca function $u=\Phi/r$ exhibits a nicely localized density. Its
derivative $-u'$ is less instructive.  Also the energy-momentum trace
of the Proca field
\begin{align}
t := -\h(\v^\a \w \Si_\a) = m^2 \Phi^2/f^2 = m^2 u^2/h^2
\end{align}
(see (\ref{energymomentum})) is quite localized. We also display its
spatial integral
\begin{align}
s(r):= 4\pi \int_r^\infty r^2 g(r) t(r)\, dr \quad \text{satisfying} \quad
\int_{\rm spatial} \v^\a \w \Si_\a = - s(0)\, dt \;.
\end{align}

\begin{figure}[t]
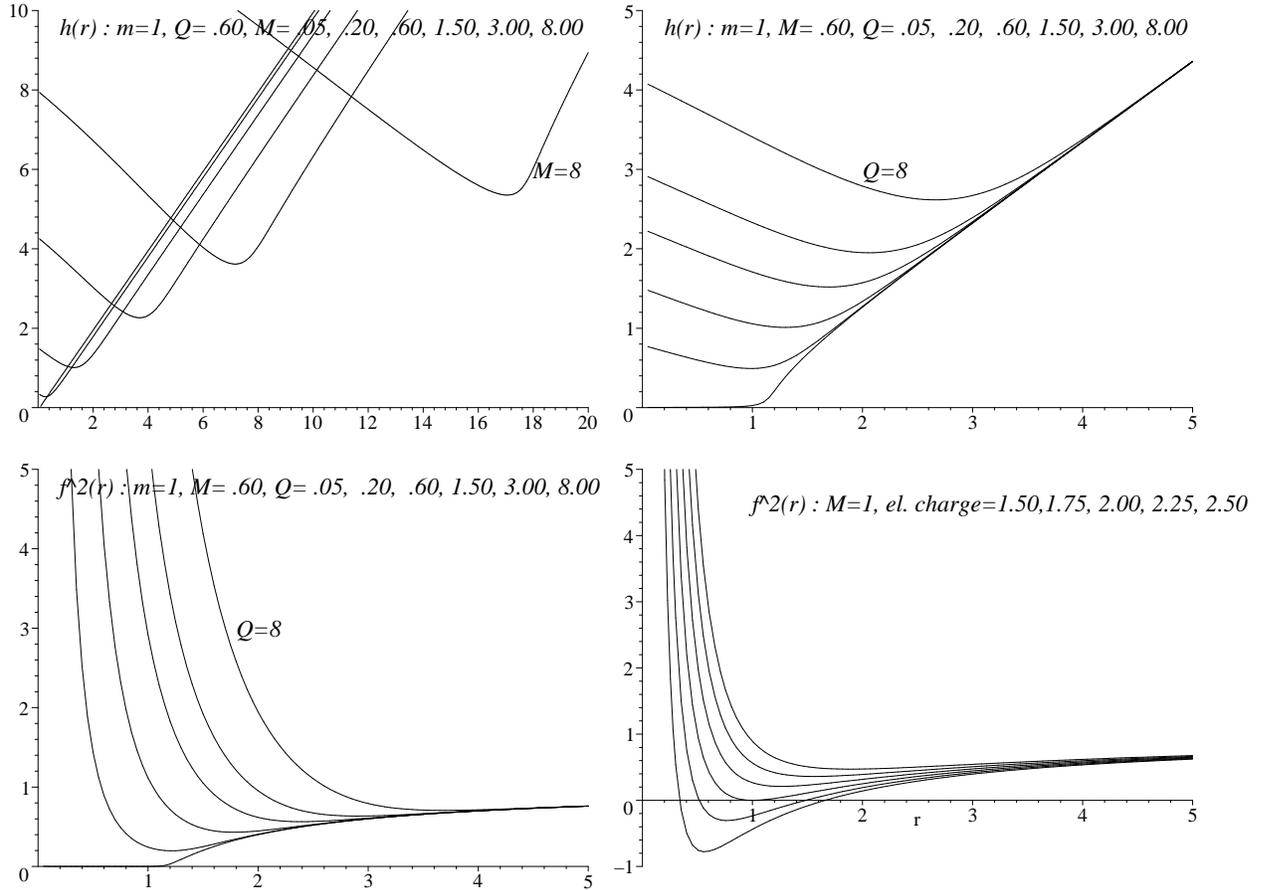
\center
\plot{hM}{0.4}
\plot{hQ}{0.4}\\
\plot{f2Q}{0.4}
\plot{reif2Q}{0.4}
\caption{
  Integration from $r_\infty$. The metric functions $h=r\,f$ and $f^2$
  are displayed for various $M$ and $Q$. For a better comparison, we
  plot the Reissner-Nordstr\"om solution in the right bottom.}
\Label{Thandf}
\end{figure}
Next we vary $M$ and $Q$. Figure \ref{Thandf} displays the metric
functions $h=rf$ and $f^2$. The most interesting point of these plots
is the following. For the Schwarzschild solution, $f^2$ becomes
negative within the horizon and vanishes at the horizon. Looking at
the energy-momentum of the Proca field $t := m^2 \Phi^2/f^2$, we can
already follow that in our case $f^2$ may neither vanish nor be
negative, as long as $\Phi$ is finite.  Hence, a Proca solution with
finite $\Phi$ \emph{may not have a horizon}. For a better comparison,
the lower two plots in figure \ref{Thandf} display $f^2$ for our Proca
system (left plot) and the square metric function of the
Reissner-Nordstr\"om solution (right plot) for varying Proca charge
and electric charge, respectively. The Reissner-Nordstr\"om solution
lacks a horizon as long as we choose the electric charge larger then
$2M$, i.e.\ the \emph{over extreme} case.  For smaller charges the
Reissner-Nordstr\"om solution has a horizon and the square metric
function becomes negative. For our Proca system the behavior is
similar for large $Q$. For smaller $Q$ though, $f^2$ \emph{approaches}
zero but never becomes negative. The important result that our
solution has no horizon is consistent with the analysis of
Ay\'on-Beato et al.\ \cite{ayonGMQ}. They proved that a static
Einstein-Proca solution may not have a horizon by considering a
spatial integral of the Proca equation. They called this a
\emph{no-hair} theorem for static black holes in the Einstein-Proca
theory -- or in the equivalent triplet ansatz of MAG. The lack of a
horizon also means that our solutions have no continuous limit for $m
\to 0$ ($m=0$ means a Reissner-Nordstr\"om solution) or $Q \to 0$
($Q=0$ generates a Schwarzschild solution)!

\begin{figure}[t]
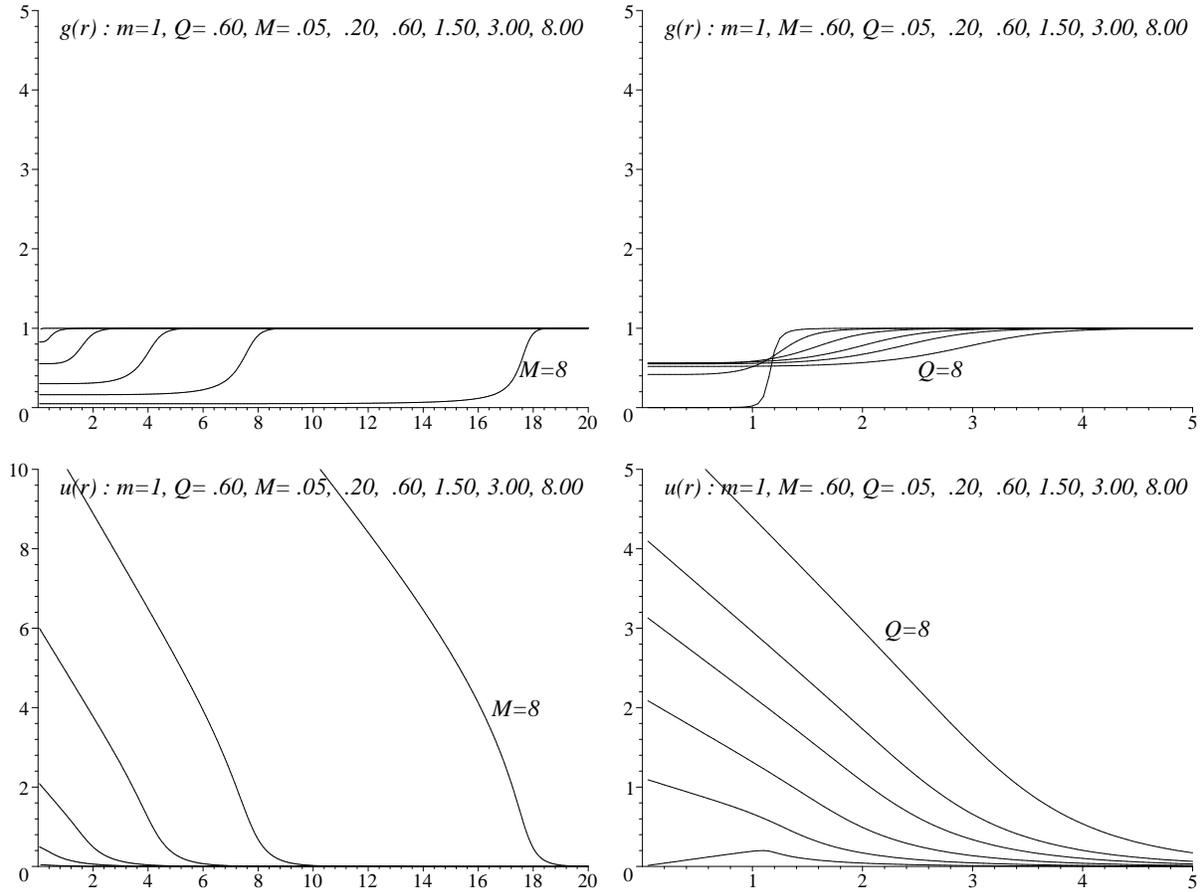
\center
\plot{gM}{0.4}
\plot{gQ}{0.4}\\
\plot{uM}{0.4}
\plot{uQ}{0.4}
\caption{
  Integration from $r_\infty$. The metric function $g$ and the Proca
  function $u$ are displayed for various $M$ and $Q$.}
\Label{Tgandu}
\end{figure}
In figure \ref{Tgandu} we plot the metric function $g$ and the Proca
function $u$. As we vary $M$, the metric function $g$ seems to
distinguish inside and outside regions. \emph{Inside}, $g$ takes some
constant value within $[0,1]$ decreasing with increasing $M$.
\emph{Outside}, $g$ equals 1. As we vary $Q$, we find that larger
values for $Q$ smear this boundary between inside and outside. Looking
at the Proca function $u$, as we vary $M$, we find that the Proca
field becomes perceptibly non-vanishing exactly within the same
boundary $g$ exhibits. Very interesting is the curve for $M=0.6$ and
$Q=0.05$ in the right plots. The Proca field vanishes as $r \to 0$ and
its derivative $u'$ becomes positive. The metric function $g$
approaches zero within the boundary instead of continuously
approaching a finite $g(0)$ as it does for larger $Q$. This behavior
is different indeed and belongs to \emph{region II} as we will explain
in the following section.

\subsection{Comparing \emph{internal} and \emph{external} parameters}

After the explicit presentation of the spherically symmetric solution
of the Einstein-Proca system, we want to examine how the external
parameters $M$ and $Q$ are correlated to the internal parameters $q$,
$b$, and $c$. Both, $Q$ and $q$, are in analogy to the Proca charge --
but with respect to different limits $r\to\infty$ and $r\to 0$,
respectively. How are they related? The computational power of Maple
allows to integrate the system for a quite large array of values of
$M$ and $Q$. For this array we calculated the values of the internal
parameters $q=u(0)$, $b=u'(0)$, and $c=g(0)$ and display them in figure
\ref{Tatzero}.

\begin{figure}
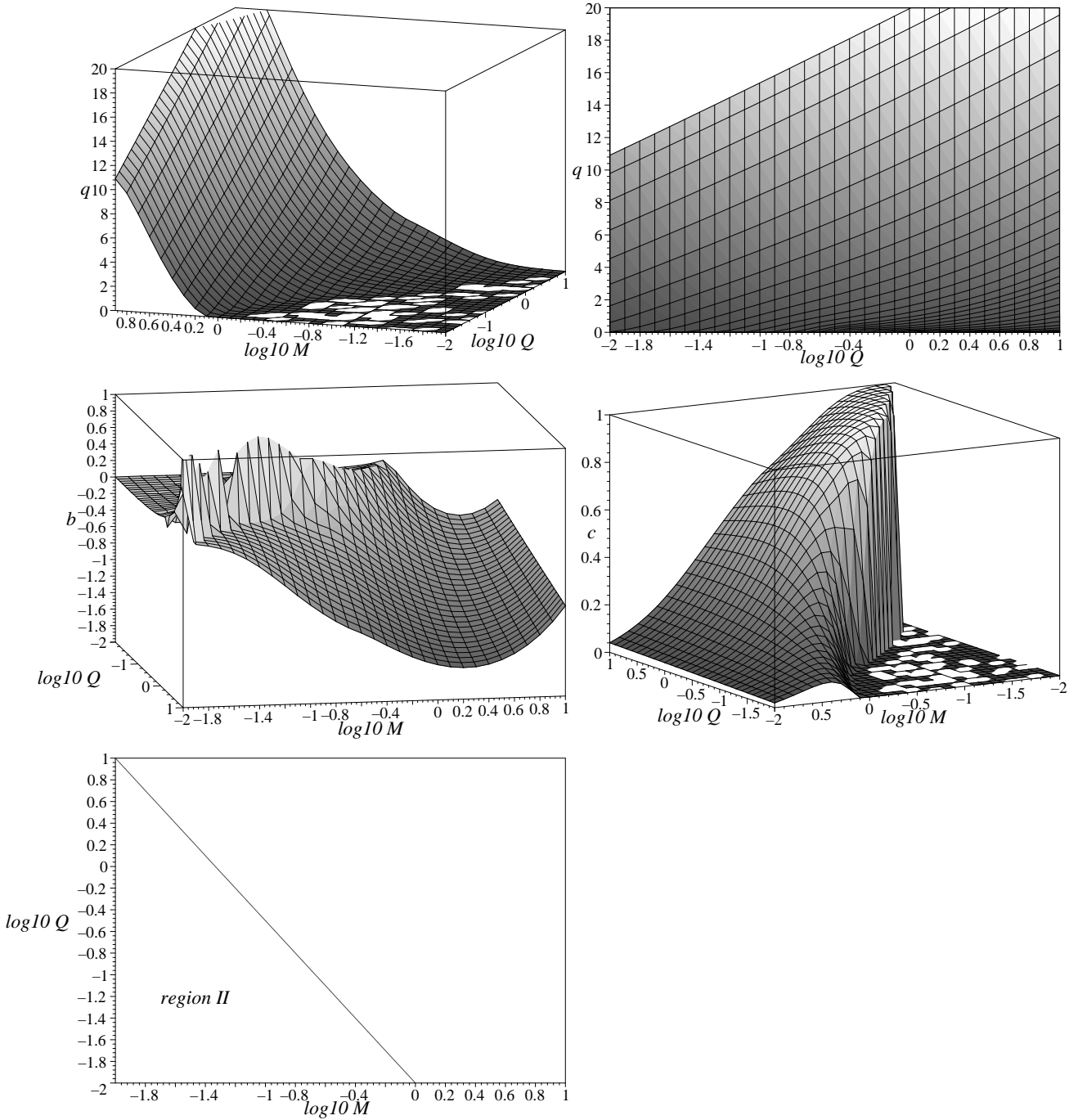

\plot{qQM1}{0.4}
\plot{qQM2}{0.4}\\
\plot{bQM1}{0.4}
\plot{cQM1}{0.4}\\
%\plot{sQM1}{0.4}
\plot{region2}{0.4}
\caption{
  The field configuration $q=u(0)$, $b=u'(0)$, and $c=g(0)$ at zero is
  displayed for an array $[-2<\log_{10} M<1,-2<\log_{10} Q<1]$ of
  different values for the external parameters $M$ and $Q$.}
\Label{Tatzero}
\end{figure}
The first two of these plots display the internal Proca charge $q$.
One can see that for any $M$, the internal $q$ depends approximately
linear on $\log_{10} Q$:
\begin{align}
q=\a \, \log_{10} Q + \b\;, \qquad \text{where roughly } 3 < \a < 4.5 \;.
\Label{qlinearlnQ}
\end{align}
In the white regions of the left plot, the numerical integration could
not reach the requested accuracy of $10^{-13}$ (relative and absolute
error). The next two plots in figure \ref{Tatzero} show $b$ and $c$.
The noisy peaks in the plot of $b$ are at the very edge to regions
where the integration could not reach the requested accuracy.
However, we observe a smooth transition to positive values of $b$.
This \emph{region II} belongs to small values of $M$ and $Q$ as the
diagram at the bottom illustrates. Also the plot of $c$ clearly
demonstrates this edge to region II but at the same time exhibits the
smooth transition to this region for smaller Q. We cannot completely
exclude that this behavior is an artifact of the numerical
integration. It is interesting that this region represents the limit
$Q \to 0$ which, as we discussed above, cannot be continuous.

\subsection{Solutions with fixed internal Proca charge}\Label{Cfixedq}

\begin{figure}[t]
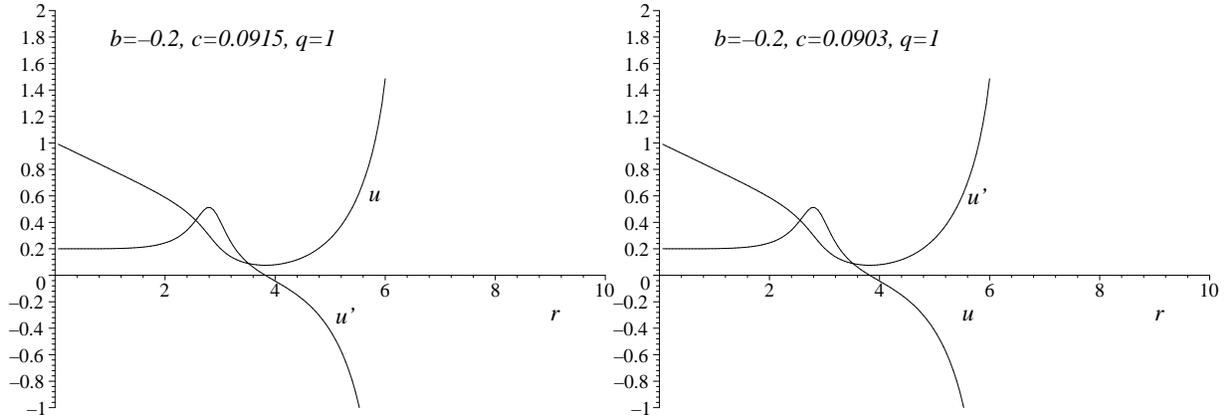

\plot{fromzero1}{0.4}
\plot{fromzero2}{0.4}
\caption{
  The integration from zero for arbitrary constants $q$, $b$, and $c$:
  If $c$ is too large (left plot) the Proca function $u$ diverges to
  $+\infty$, whereas if $c$ is too small (right plot) $u$ diverges to
  $-\infty$.  Note the small difference of 0.0012 between the two
  values of $c$.}
\Label{Tfromzero}
\end{figure}
As we already mentioned, the integration from zero is quite costly. If
we start integration with the constants (\ref{zeroconst}) with
arbitrary $q$, $b$, and $c$, the solution diverges at some finite
radius. Figure \ref{Tfromzero} displays the two possible divergences:
the Proca function $u$ either diverges to $+\infty$, if $c$ is large,
or to $-\infty$, if $c$ is small. Only a fine tuning of $c$ allows to
find a global solution by an integration from zero. Of course, these
integrations exhibit the same solution as an integration from
$\infty$. Since this procedure is very time expensive, we had to find
another way to produce solutions with fixed internal Proca charge by
using the relation (\ref{qlinearlnQ}) between $q$ and $Q$. We fixed
$q$ by tuning $Q$ for given $M$. This may be done very quickly because
if we have found one solution with arbitrary $q$, relation
(\ref{qlinearlnQ}) tells us of how to approximately choose $Q$ for a
given value of $q$.  Thereby we need only about five steps to fix $q$
on a given value up to an accuracy of $10^{-5}$. Figure \ref{TtuneQ}
shows how $Q$ has to be chosen for different $M$ in order to fix
$q=1$. Figure \ref{Tfixedq} displays the solutions for fixed $q=1$ and
different $M$. The plot of $h$ nicely demonstrates the necessary
relation $h(0)=q/\sqrt{2}$ we found in (\ref{zeroconst}).  The plot of
$f^2$, again, demonstrates that our solutions \emph{do not have a
  horizon}. Instead, $f^2$ approaches zero but never becomes negative.
Finally, the plots of $g$ and $u$ exhibit the localization of our
Proca particle within a finite radius.

\begin{figure}[t]
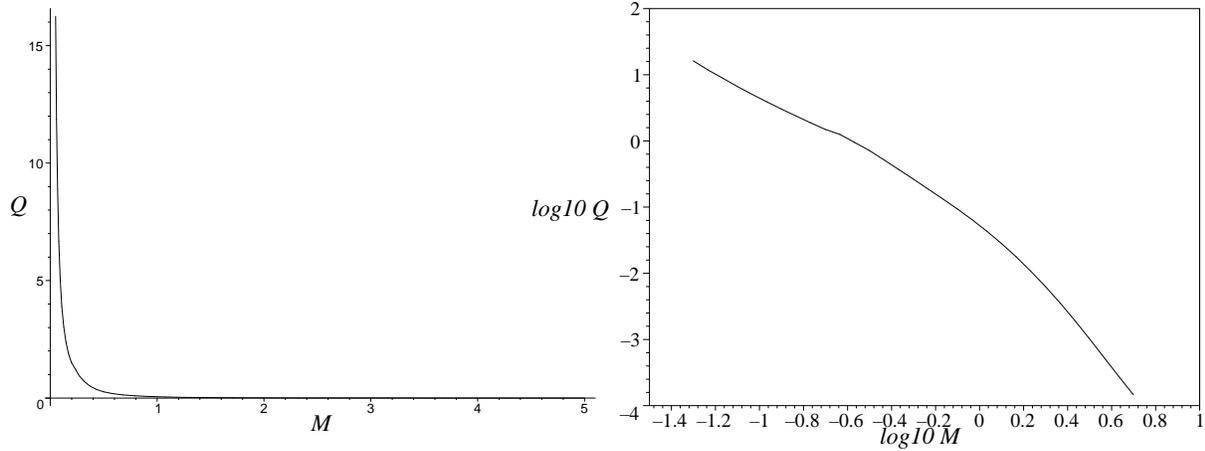

\plot{QMq1}{0.4}
\plot{QMq2}{0.4}
\caption{
  How to choose $Q$ for given $M$ in order to fix $q=1$.}
\Label{TtuneQ}
\end{figure}

\begin{figure}[t]
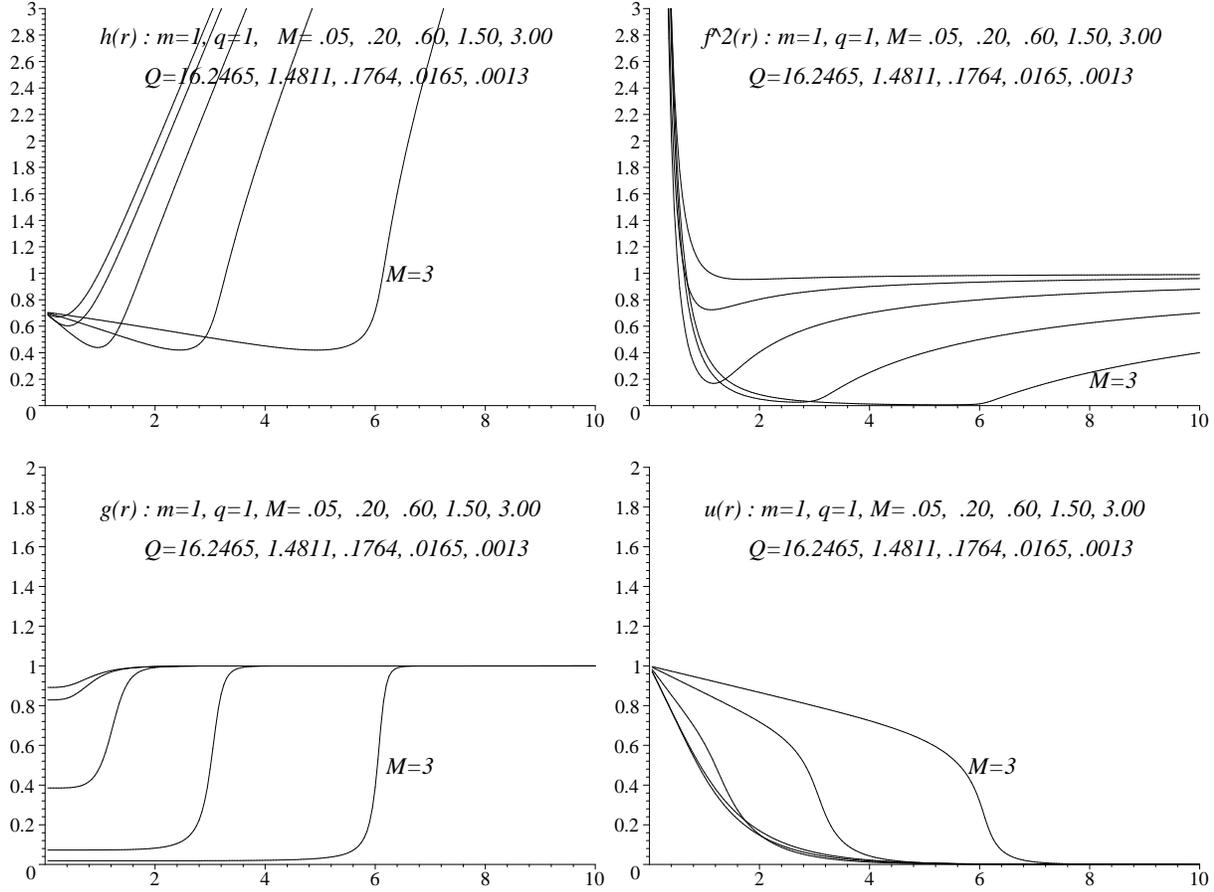

\plot{hMq}{0.4}
\plot{f2Mq}{0.4}\\
\plot{gMq}{0.4}
\plot{uMq}{0.4}
\caption{
  The metric functions $h=r\,f$, $f^2$, and $g$ and the Proca function
  $u$ for fixed internal Proca charge $q=1$.}
\Label{Tfixedq}
\end{figure}

\section{Other approaches}

\subsubsection*{Failure of the magnetic type ansatz}

In section \ref{Cflat} we found a solution $\phi =
p\,\exp(-mr)\,(1-\cos\t)\, d\p$ for the flat Proca equation
(\ref{procaeq}). However, with the general spherically symmetric
ansatz (\ref{coframe}) for the coframe, we find the 12-component of
the Einstein equation:
\begin{align}
\v^1 \w X^2 = \frac{f \k u u' (\cos\t-1)}{g r^3 \sin\t}\ \eta
\end{align}
which has only trivial solutions. Hence, there exists no magnetic
analogue to the previous solution!

\subsubsection*{Rosen's ansatz}

In this section we have a glance on the ansatz of Rosen \cite{rosen}.
He considered the lagrangian (\ref{procalag}) of a Proca 1-form
together with the ansatz
\begin{align}
\phi = w_0(r) {\rm e}^{-i \o t}\, dt + w_1(r) {\rm e}^{-i \o t}\, dr \period
\Label{ros1}
\end{align}
With (\ref{ros1}) and the spherically symmetric coframe (\ref{coframe}),
the Proca equation (\ref{procaeq}) reads
\begin{align}
0 &= \v^0\w\v^2\w\v^3 ~ \frac{{\rm e}^{-i \o t}}{fg} ~
     \Big[ w_1(f^2 m^2 - \o^2) - i \o w_0' \Big] \feed
&- \v^1\w\v^2\w\v^3 ~ \frac{f {\rm e}^{-i \o t}}{g^2 r} ~ 
   \Big[ 2 i \o w_1 + g^2 m^2 r w_o /f^2 -2 w_0'\feed
&  \hspace{4cm}  - (w_0'' - i \o w_1') r + (w_0' - i\o w_1)r g'/g \Big] \period
\Label{ros3}
\end{align}
From the 023-component we read off
\begin{align}
w_1 = \frac{i \o w_0'}{f^2 m^2 - \o^2} \;,
\end{align}
which is equivalent to equation \cite{rosen} (17). We substitute $w_1$
and identify $w \equiv w_0$. \emph{In flat space}, the 123-component
of the Proca equation (\ref{ros3}) becomes
\begin{align}
0 = r w'' + 2 w' + (\o^2 - m^2) r w \comma
\Label{ros2}
\end{align}
which is in agreement with \cite{rosen} (19). Comparing with
(\ref{procaflateq}), we find that this equation is the same as the
Proca equation in flat space for a Proca 1-form $w(r)\, dt$ with mass
parameter $\sqrt{m^2 - \o^2}$. Hence, if we set $\o = m$, as Rosen
proposes in equation \cite{rosen} (33) (in his notation $C=1
\Rightarrow \o=\kappa$), then (\ref{ros2}) is the ordinary,
\emph{massless} Maxwell equation. Thus, in flat spacetime, such a
particle has no finite extension. This raises the question of how to
choose the initial constraints at infinity for such a numerical
integration of the field equations. Finally, Rosen assumes that the
total gravitating mass ($M$ in our notation) is equal to the mass
parameter $m$ (in dimensionless units). In general, one can hardly
compare Rosen's work with ours because he concentrates on the idea of
an \emph{elementary particle with finite and absolute boundary}
existing in the Einstein-Proca theory. Thus, he assumes an
\emph{empty} (exactly Schwarzschild) space outside the particle's
boundary -- and not a space that becomes Schwarzschild asymptotically,
as we did. He calculates the solution by continuously (not smoothly)
fitting the (Einstein-Proca) fields inside to the (purely Einstein)
fields outside.

\section{Summary}\Label{Cprocadisc}

The introduction of this chapter explained the meaning of the coupled
Einstein-Proca theory as an effective theory of MAG and thus motivated
our analysis of this theory. Most interesting, we found the general
condition (\ref{zeromass}) for the massless case, i.e.\ for the
(restricted) lagrangian being equivalent to the Einstein-Maxwell
theory. Then we derived the field equations (\ref{procaeq},
\ref{einsteineq}) and the energy-momentum (\ref{procaem}) of the
Einstein-Proca theory and displayed the (electric  and magnetic type)
Yukawa solution in \emph{flat} spacetime. For an electric type ansatz
we discussed the numerical integration and its integration constants
and also offered the power series expansion (\ref{power}) at the
origin. We also proved the failure of the magnetic type ansatz. Here,
we collect the essential features of the numeric solution:

(1) In figure \ref{Tall} we display the typical solution of the
Einstein-Proca system for the case of the gravitating mass $M$ and the
external Proca charge $Q$ being of the same order as the mass
parameter $m$.

(2) Figure \ref{Thandf} concentrates on the behavior of the metric
function $f$, with $\v^\0=f\, dt$. We found that our solution has
\emph{no horizon}, which should also be clear from the energy-momentum
trace in (\ref{energymomentum}) and is consistent with \cite{ayonGMQ}.
Hence, our solution has a naked singularity. The lack of a horizon
also prohibits a continuous limit to the Reissner-Nordstr\"om ($m \to
0$) or Schwarzschild ($Q \to 0$) solution.

(3) Figure \ref{Tgandu} focuses on the shape of the Proca particle. We
found some boundary which is sharp for small external Proca charge
$Q$.  The larger the gravitating mass $M$, the larger the extension of
the Proca particle.

(4) Figure \ref{Tatzero} exhibits the interesting linear relation
(\ref{qlinearlnQ}) between the internal Proca charge $q$ and the
logarithm of the external Proca charge $\log_{10} Q$.

(5) The $b$- and $c$-plots in figure \ref{Tatzero} and the $g$- and
$u$-plots in figure \ref{Tgandu} suggest a different kind of behavior
for small $M$ and $Q$ (region II). Note that this region represents
the limit $Q \to 0$. Although the transition to this behavior is
smooth, we cannot completely exclude it to be an artifact of the
numerical integration.

\subsection*{Acknowledgment}

The author is grateful to Prof.\ Friedrich W. Hehl (University of
Cologne) and Yuri Obukhov (Moskow State University) for their support.

\end{document}